\documentclass[11pt,dvips,twoside,letterpaper]{article}
\usepackage{graphicx}
\usepackage{times}
\usepackage{fancyhdr}
\usepackage{geometry}
\usepackage{latexsym}

\begin{document}
\title{
~\\[-1.2in]
{\normalsize\noindent
\begin{picture}(0,0)(208,-3.15)
\begin{tabular}{l p{0.60in} r}
In: Trends in Quantum Gravity Research   & & ISBN 1-59454-670-3\\
Editor: David C. Moore, pp. 109-137  & & \copyright 2006 Nova Science Publishers, Inc.\\
\end{tabular}
\end{picture}}
{\begin{flushleft} ~\\[0.63in]{\normalsize\bfseries\textit{Chapter~3}} \end{flushleft} ~\\[0.13in] 
\bfseries\scshape Quantum Mechanics, Quantum Gravity,\\ 
and Approximate Lorentz Invariance\\
from a Classical Phase-Boundary Universe}}
\author{
\bfseries\itshape Michael Grady\\
Department of Physics, SUNY Fredonia,
Fredonia NY 14063 USA }
\date{}
\maketitle
\thispagestyle{empty}
\setcounter{page}{109}

\pagestyle{fancy}
\fancyhead{}
\fancyhead[EC]{Michael Grady}
\fancyhead[EL,OR]{\thepage}
\fancyhead[OC]{Quantum Mechanics, Quantum Gravity, 
and Approximate Lorentz Invariance...}
\fancyfoot{}
\renewcommand\headrulewidth{0.5pt}
\addtolength{\headheight}{2pt} 
\headsep=9pt

\begin{abstract}
A classical dynamical system in a four-dimensional Euclidean 
space with universal time is considered. 
The space is hypothesized to 
be originally occupied by a uniform 
substance, pictured as a liquid,
which at some time became supercooled.
Our universe began as a nucleation event initiating
a liquid to solid transition. The universe we
inhabit and are directly aware of consists of only the 
three-dimensional
expanding phase boundary - a crystalline surface. Random energy
transfers to the boundary from thermal fluctuations
in the adjacent bulk phases are interpreted by us as quantum
fluctuations, and give a physical realization to the stochastic
quantization technique.  
Fermionic matter is modeled as screw dislocations;
gauge bosons as surface acoustic waves.  Minkowski space emerges
dynamically through redefining local time to be proportional
to the spatial coordinate 
perpendicular to the boundary. Lorentz invariance is only 
approximate, and
the photon spectrum (now a phonon spectrum) has a maximum energy. 
Other features include
a geometrical quantum gravitational theory
based on elasticity theory,
and a simple explanation of the quantum measurement process 
as a spontaneous symmetry breaking.
Present, past and future are physically distinct regions, 
the present being a 
unique surface where our universe is being continually constructed.
\end{abstract} 


\section{Introduction}

In the following, a new picture of the big bang and the underlying
structure of the universe is proposed, based on a classical field
theory in a four-dimensional Euclidean space with a universal time
(a 4+1 dimensional theory)\cite{grady98,rubakov,akama,otherdomain}.
The big bang is treated as a 
nucleation event
for a first-order phase transition (pictured as a liquid to 
solid transition)
and our universe is the three-dimensional phase boundary 
between the expanding
solid and preexisting liquid phases. This classical brane-theory 
appears to have the potential
to explain a diverse set of phenomena 
--  Lorentz invariance, quantum fluctuations and zero-point energy, 
quantum superposition and measurement, 
elementary fermions and bosons, gauge forces, gravity, 
the big bang and a non-decelerating
expansion of the universe.
It is possibly rich enough to 
give a ``theory of everything" from a relatively simple base-theory 
consisting of a small
number of elementary atoms or molecules 
and basic elastic forces holding 
them together.
In this model, all of the forces and particles of standard particle 
theory are
secondary effects, consisting of the collective excitations and 
dislocations of the base-theory, 
just as in condensed matter physics where such 
excitations play a pivotal role,
reducing the elementary degrees of freedom to a mere background 
for the more interesting
and important collective excitations.

We begin by assuming a four-dimensional
Euclidean space, filled with a uniform fluid at some temperature,
undergoing thermal fluctuations. In addition to the four spatial
dimensions, there is also a universal time.  Another possibility would
be to start already with a five-dimensional Minkowski space, however
this does not seem to be necessary.
This liquid was cooling, 
became supercooled, and at some point a solid crystal nucleated.
This was the big bang. The universe begins as a 
fluctuation, already at a finite size, because in order to grow
rather than shrink, the initial crystal must be large enough
that the positive surface energy is less than the negative 
volume energy
relative to the liquid. 
In such a model there is no physical singularity at the beginning
and there is no reason for the universe to be particularly hot or dense
at this time either (more on this later).
The {\em surface} of the solid, 
the phase boundary, is
an expanding three-dimensional space, our universe. 
This differs from other ``bubble universe'' pictures, where the 
universe is the {\em interior} of a {\em 3-d} bubble.
In fact, it bears an uncanny resemblance to the 
simple ``expanding balloon"
model which is often used as an example of a uniformly expanding
curved space. However the present model differs in that 
the interior and exterior of the 
balloon are real spaces, though not directly observed by us.
We are directly aware only 
of the phase boundary separating the phases,
which we refer to as the ``present''.  As the
crystal grows, this hypersurface, 
our universe, expands. Already there
is a variance with the usual $\Lambda=0$ Friedmann universes.
Namely, our universe is closed, but will expand forever. The pressure
on the surface caused by the energy difference of the two
phases acts something  
like a repulsive cosmological constant. 
This universe actually expands faster 
as time goes on, not slower. If, as
is likely, dissipation is present, it will eventually 
approach a constant rate. (This assumes a constant amount 
of supercooling -- if the 
base liquid cools 
more, the 
expansion rate could continue to 
increase as the degree of supercooling 
increases. Without dissipation, the expansion rate
increases exponentially).
Recent astrophysical evidence shows that the expansion rate is not
slowing, 
but may even be speeding up\cite{expand} which is consistent 
with this scenario. 

In the following, the emergence of a quantum field theory on the 
surface and the origin of quantum fluctuations is 
discussed in section two.
The relation between real and imaginary time path integrals 
is clarified as a difference between 
non-equilibrium and equilibrium statistical mechanics.
Section three deals with the dynamical realization of Lorentz invariance 
and special relativity, including possible tests of the theory,
and consequences for cosmic ray physics.
In Section four, the description of photons as surface acoustic waves is
explored. The Plank relation, $E=h \nu$, and zero point energy 
are derived, with Plank's constant being 
essentially the four-dimensional temperature. 
Section five describes the interpretation and realization
of quantum superpositions and quantum measurements.
Section six discusses four-dimensional dislocations as candidates for 
elementary fermions. 
The possibility of modeling quarks as 
partial dislocations, which, in ordinary crystals are 
naturally confined,
is explored.
Section seven outlines the likely gravitational theory that 
results from 
the relationship between the curvature of the surface and 
the presence of
dislocations and interstitials, following previous analogies 
drawn by many 
authors between elasticity theory and general relativity. Modeling
fermions as screw dislocations introduces a natural relation between
spin and torsion, as in the Einstein-Cartan theory, which may be a good
continuum approximation to the underlying lattice theory. 
Whatever gravitational theory that results is 
automatically a quantum theory of gravity since the 4-D 
thermal fluctuations
are present in the surface. 
Section eight discusses the cosmology of the model, including possible 
difficulties in fitting observations.
Section nine discusses the rather different nature of time 
that the model
presents and relates it to Whitehead's conception of time. 
The different causality 
structure due to the model having a preferred
frame is discussed 
(special relativity is only approximately realized).

\section{Quantum Field Theory from a Classical Field Theory}

\enlargethispage{12pt}
The basic theory needed to describe this 
expanding phase boundary is non-equilibrium classical 
statistical mechanics. The boundary itself may be 
describable in terms of dynamical critical phenomena\cite{dcp}.
The solid, in some sense, lies in the past, since
we have been there earlier, although it still exists 
in the present when
observed from the higher dimension.  The liquid represents
the future, since that is where we are going, but it also exists
now, as an undifferentiated, fluctuating medium. To distinguish the 
current states of the solid and the liquid from our own 
past and future,
they may be called the ``current past'' and ``current future''. 
They differ from our past and future because changes 
may have occurred
after the solid was formed, and the future certainly will be 
different when we
arrive there.
To the extent that the solid is frozen, however,
our past may be accurately preserved within it.
We may not be aware of the existence of the 
liquid due to its uniformity.
However, the boundary which we inhabit is in thermal contact
with both the liquid and solid phases, and can 
certainly exchange energy
with them. 
Actually, since the surface is continuously colonizing 
new parts of the liquid, the mountain, in this case, 
is moving to Mohammed.
Energy fluctuations that were present in the adjacent liquid will be
incorporated into the ``new surface" an instant later. 
These will interact
with propagating surface modes 
which are passed from the ``old surface" to
the ``new surface" as each layer is added.
Thus waves riding the interface
will experience random energy
fluctuations from this thermal contact. These 
random 4-d thermal fluctuations
could explain quantum fluctuations. 

It is well known
that in ordinary quantum theory, if Minkowski 
space is analytically continued
to Euclidean space, quantum fluctuations behave  
as higher-dimensional thermal fluctuations, i.e. the Feynman
path integral becomes an ordinary statistical mechanical 
partition function
in 4 (+1) dimensions (in equilibrium statistical mechanics there is
an implied time dimension).
Plank's constant is proportional to the temperature of 
the four-dimensional
Euclidean space. The existence of a 1-1 mapping between quantum field 
theory and statistical mechanics in one more dimension opens the 
possibility that the physical reality that quantum theories are 
describing actually corresponds to a higher dimensional classical 
theory, one for which, if all degrees of freedom were accounted for, 
would constitute a dynamical system of some kind. 
Aside from the important new feature of an extra dimension,
this is essentially
the point of view of Nelson\cite{nelson}, 
whose stochastic quantization technique
attempts to explain the fluctuations of quantum mechanics through 
interaction with an otherwise unobservable 
fluctuating background field. Stochastic 
quantization was extended to field theory by Parisi 
and Wu\cite{parisiwu}, who
showed the equivalence of the Euclidean path integral to a stochastic
process controlled by a Langevin equation, which operated in a 
fictitious new time, completely unrelated to ordinary time. 
Whereas this can be seen as simply a mathematical tool, some have
speculated that the reformulation could be closer to 
reality. Of course,
to the extent that mathematical formulations are equivalent, it does
not really matter to the physicist which is ``more real", however if
our current theories are only approximations, 
then such considerations
make sense in trying to find a more correct and 
accurate theory. If
a stochastic differential equation explains quantum 
fluctuations, then
this would likely be the case, since in most cases one can picture
such equations as approximations 
resulting from more detailed deterministic
dynamical systems for which some degrees of freedom have been 
averaged over.

The main problem in making sense of this connection between quantum 
systems and classical systems in one higher dimension is the analytic
continuation to imaginary time, and the lack of any apparent connection
between the ``Langevin time" of a Langevin simulation and real time.
However, if one considers the behavior of fields that live on an
expanding phase boundary in a 4-d Euclidean space,
such a connection can be made. If one accepts the Langevin time 
itself as real time, then there will be a connection 
between it and the 
fourth spatial coordinate 
at the surface (the coordinate perpendicular
to the surface), due to the motion of the surface. For the 
sake of simplicity it will be assumed to travel at constant speed. 
For observers riding the surface, the fourth spatial coordinate 
will be nearly indistinguishable 
from time, since they increase in lockstep.
In a following section it will be argued that this identification
leads to a ``spatialization" of time from which all of the properties
of special relativity arise - in particular it will be 
seen that clocks
constructed from dislocations and surface modes do not 
keep universal time,
but rather the local time of special relativity.

The remaining question 
is why quantum field theory is given in terms of
a real-time path 
integral with an oscillating exponential rather than 
the imaginary-time version with a real exponential. It is perhaps not
a question of real or imaginary time which is a mathematical 
transformation with no apparent physical basis, but the rather less
exotic notion of real vs. imaginary frequency 
describing oscillatory vs. overdamped motions.
This can also be seen as the difference between non-equilibrium 
and equilibrium statistical mechanics. If the universe were a single
phase in equilibrium then it could be described by an equilibrium
statistical mechanical ensemble. Correlation functions would be 
decaying real exponentials. The corresponding Langevin equation would
be dominated by dissipative forces and the corresponding path integral
would be Euclidean (i.e. the imaginary time version). 
However, an expanding phase boundary is a decidedly non-equilibrium 
object. It breaks time translation invariance
and at least one spatial translational invariance.
One may also have propagating modes present on the surface, due to 
conservation laws. Such propagating modes exhibit oscillatory rather
than dissipative behavior, and occur in many 3-d systems\cite{dcp}. 
They lead
to various complications in the theory of dynamical 
critical phenomena,
and are a crucial feature in the dynamical theory of 
phase transitions.
In many cases these systems are still describable by a stochastic
differential equation - a {\em complex} Langevin equation, where
non-dissipative forces play a crucial role\cite{dcp,haken}. 
Solutions are oscillating
but contain random phase and amplitude fluctuations. 
The Fourier transforms of correlation functions contain real-axis 
poles.

A number of
authors have shown that the Parisi-Wu stochastic quantization can 
be performed directly in Minkowski space, the result being a complex
Langevin equation which will be exhibited 
shortly\cite{minkpi,othermink}. 
This completes
the logical connection. To sum, fields which represent dislocations
or collective modes on a moving phase boundary in a 4-d Euclidean
space are likely describable by a complex Langevin equation, which
approximates the behavior of the larger deterministic 
dynamical system which
fills the entire Euclidean space. This complex Langevin equation
has an equivalent path-integral representation (meaning the two systems
have the same correlation functions), which resembles the Minkowski
space path integral of quantum field theory. Some details will likely
be different, however. For instance, it does not seem likely that
dissipation will be entirely absent from the surface. This could be 
countered by an energy input, resulting in a steady-state rather than
an isolated system. Such a system lacks time-reversal invariance at 
some level, which could have observable consequences (and perhaps
help to explain CP non-conservation in the $K^0 -\bar{K}^0$ system).

The Langevin equation is a first-order differential equation with
a fluctuating random force. It was first applied to the case of 
Brownian motion of a small particle in a background of randomly
moving molecules colliding with it. If $v$ represents the velocity
of the particle, then the Langevin equation is 
\begin{equation}
\dot{v}= -\gamma v +F +	\eta(t)
\end{equation}
The $\gamma v$ term is the frictional force exerted by the fluid,
$F$ is an applied external force (if present) such as an electric 
field,
and $\eta(t)$ is the fluctuating force designed to mimic the many 
collisions between the fluid which is assumed to be in thermodynamic
equilibrium at some temperature and the particle. In the absence of
force $F$, the particle exhibits a random walk in position. Without
the damping term it would also perform a random walk in velocity, and
the kinetic energy would increase without bound. However, any amount
of dissipation is sufficient to stabilize it and the particle's average
kinetic energy will become equal to $\frac{1}{2} {\rm k} T d$,
where $d$ is
the number of spatial dimensions, $T$ is the fluid temperature, and
k is Boltzmann's constant.
If one wants to extend this treatment to an oscillator, a problem
arises in that a position dependent force cannot be incorporated into
a first order equation. 
The Hamilton equations are, of course, first order, 
but there are two of
them. By introducing a complex variable $b=(p+ix)/\sqrt{2}$, 
$b^* = (p-ix)/\sqrt{2}$,
one can write the Hamilton equations 
for one degree of freedom as a single complex equation:
\begin{equation}
\dot{b} = i \frac{\partial H}{\partial b^*} .
\label{eqn2}
\end{equation}

To explore these ideas in more detail, consider the system of
a single harmonic oscillator interacting with a bath of other harmonic
oscillators\cite{haken}. 
The simple harmonic oscillator in coordinates 
$x=\sqrt{m \omega } x'$, $p=p'/\sqrt{m \omega }$, where $x'$ and $p'$ 
are the usual coordinate and momentum has the Hamiltonian 
\begin{equation}
H=\omega b^* b
\end{equation}
Here, $k$ is the spring constant and $\omega \equiv \sqrt{k/m}$. 
The Hamilton equation (\ref{eqn2}) becomes
\begin{equation}
\dot{b}= i \omega b .
\end{equation}
Interestingly, this formalism can be easily extended to the damped
oscillator\cite{haken,dekker} 
by allowing $\omega$ to become complex. Replacing $\omega$
with $\omega + i \gamma$ gives the equation of motion for the damped
oscillator,
\begin{equation}
\dot{b} = i \omega b - \gamma b.
\end{equation}
Here, the complex formalism goes beyond the real formalism, since the 
Hamiltonian does not technically exist for the damped oscillator
unless auxiliary fields are added\cite{dekker}. Note
that this is not a fully complex Hamiltonian function which 
would result
in doubling the number of equations of motion and producing an 
overdetermined system. Rather, the Hamiltonian takes values along a ray
other than the real axis.
If we add a fluctuating force, one obtains a complex Langevin equation,
\begin{equation}
\dot{b} = i \omega b - \gamma b +\eta (t). \label{eqn6}
\end{equation}
This equation can be derived as the equation of motion of a tagged 
oscillator interacting with a collection of ``bath" oscillators
whose behavior is averaged over\cite{haken}. 
The bath provides both the random
force and the damping.  
It can be used, for instance, 
to describe
the behavior of a single-mode laser interacting with a 
thermal medium
and thermal mirror fluctuations\cite{haken,laser}. 
Similarly, it can also be used to describe 
propagating modes in dynamical critical phenomena\cite{dcp}. 
Thus the complex
Langevin equation is a well-established equation for describing
oscillating or propagating modes in a random medium. 

If the Parisi-Wu quantization is applied to the Minkowski field 
theory directly, it has been shown\cite{minkpi,othermink} 
that the correlation functions
derived from the path integral
\begin{equation}
\int D \phi \exp (i S(\phi )/\hbar)
\end{equation}
can be obtained from the long-time behavior of the Langevin equation
\begin{equation}
\dot{\phi} = i\delta S/\delta \phi ^{*} - \epsilon \phi + \eta(x ,t)
\label{eqn8}
\end{equation}
where $t$ is a fictitious ``Langevin time" unrelated to the real
time in the path integral, $x$ represents the four 
space-time variables, $x_i$, with $i=1..4$,
and the Gaussian noise term has the following correlation function:
\begin{equation}
<\eta^{*}(x ,t) \eta(x' ,t')>= 2 \hbar \delta ^4 (x -x')
\delta (t-t') .
\end{equation}
 Field correlations are computed at equal
Langevin times and the damping, $\epsilon$, 
is taken to zero after correlation
functions are calculated.  Equation \ref{eqn8} 
appears to be a multivariate version
of equation \ref{eqn6} (the first term being generalized to the RHS of 
equation \ref{eqn2}) with the 
Minkowski action $S$ playing the role of a 
Hamiltonian. For example, for the complex scalar field, 
\begin{equation}
S=\int (|\partial _{\mu} \phi|^2 -m^2 \phi ^* \phi) d^4 x
\end{equation}
one obtains the complex Langevin equation
\begin{equation}
\dot{\phi} = i (-\partial ^2 \phi/\partial x_4^2 + \nabla ^2 \phi 
-m^2 \phi) - \epsilon \phi + \eta (t) \label{eqn11}
\end{equation}
One can understand the difference in sign between the spatial and
local-temporal ($x_4$) 
derivative terms in relation to the different dynamic behavior
of the interface in these directions. If one thinks of $\phi$ as a 
displacement field of elementary atoms from their quiescent-crystal
locations, one expects oscillatory behavior in the spatial directions.
The sign of the $\nabla ^2$ term is such as to provide the usual 
restoring force from neighboring atoms 
making this possible. A negative restoring force, 
as exists in the time direction, leads to 
an instability, as occurs in a $\phi^4$ theory with 
a negative mass-squared term,
for example. This will result in translational motion
(a soft mode). If we think
of the membrane as the physically relevant object, it is 
in translational
motion in the temporal direction. Therefore the ``Minkowski signature"
of the D' Alembertian operator would appear to be directly related to 
the dynamics of the phase boundary, which is itself, of course, 
controlled by the Lagrangian 
of the ``base-theory'' of the elementary atoms.
The fact that the instability 
that resulted in the motion of the phase boundary
is a phase transition of the base-theory, which is likely driven by
a spontaneous symmetry breaking,
suggests that the Minkowski space we are
familiar with is due to a spontaneous symmetry breaking from original
space-time symmetry of the base-theory.
Such a dynamical origin for Minkowski space, and the consequences
of special relativity that result, is in rather distinct contrast to
the kinematical origin postulated by Einstein. Indeed it is more
like the view held by Lorentz and others who clung to the idea of 
a cosmic ether, even if invisible. The crystal and liquid in the 
picture presented here is a form of ether, which, however, is only
invisible at low energies. When photon wavelengths get close to the
elementary lattice spacings, then the deviation from linearity of
their phonon-like dispersion relations in this theory
will become apparent, and the existence of the crystal will have
observable effects. These ideas are expanded in secs. III \& IV.

Getting back to the Langevin equation under discussion,
we now consider the consequences of our somewhat 
different interpretation
of the Langevin time coordinate. The usual treatment calculates
correlation functions at equal Langevin times, whereas we are
essentially locking the Langevin time to the ordinary time through the 
presumed uniform motion of the phase boundary.
It is important to see whether this will make any difference in the
relationship to the quantum field theory.
One notices
a peculiarity in equation \ref{eqn11}
when subjected to dimensional analysis. Taking the
usual dimension of $[ \ell^{-1} ]$ for the $\phi$ field and 
$[\ell ]$ for $x_i$ and $t$ variables leads to different 
dimensions for the 
$\dot{\phi}$ and $\Box \phi$ terms. One common solution is to let the 
fictitious time have dimensions $[ \ell ^2 ]$\cite{minkpi,namiki}. 
Then dimensional consistency is obtained and $\hbar$ comes out
dimensionless. However, since we want the fictitious time to become
the real time, another solution must be taken. 
Introducing a parameter $a$ with dimensions of length, which can be
taken to be the lattice spacing, rewrite equation \ref{eqn11} as
\begin{equation}
\dot{\phi} = i a (-\partial ^2 \phi/\partial x_4^2 + \nabla ^2 \phi 
-m^2 \phi) - a \epsilon \phi + \eta (x,t)
\end{equation}
where
\begin{equation}
< \eta^{*} (x,t) \eta (x',t') > \, = 
2a \hbar \delta^4 (x-x') \delta (t-t') .
\end{equation}
The two times now have the same dimensionality, the equation is 
dimensionally consistent and
the two factors of $a$, one multiplying $S$ and one multiplying
$\hbar$ will cancel in the path integral ($\hbar$ will now 
be set to unity).
For the free field theory we are considering here, the Langevin
equation can be solved\cite{parisiwu,minkpi,namiki}, 
with a long time stationary correlation
function  
\begin{equation}
D(x-x',t-t') \equiv \lim _{t,t' \rightarrow \infty} 
<\phi ^* (x,t) \phi (x',t') >
\end{equation}
(with $t-t'$ fixed) given by
\begin{equation}
D(x-x',t-t') = \frac{2a}{(2\pi )^5} \int d^4 k \int d \omega
\frac{e^{-ik(x-x')-i\omega (t-t')}}{\omega^2 
+ a^2(k^2-m^2+i \epsilon )^2} .
\end{equation}
Setting $t-t'=x_4 - x_4 '$, we get a free propagator of 
\begin{equation}
D(x-x')= \frac{1}{(2 \pi)^4} \int d^4 k 
\frac{e^{-ik(x-x')}e^{-a|(k^2-m^2)(x_4-x_4')|}}
{k^2 - m^2 + i \epsilon } .
\end{equation}
This is slightly modified from the usual field-theory propagator
which results from taking equal Langevin times, $t=t'$.
However, the extra exponential 
factor affects only the off mass-shell propagator,
and even for that is highly suppressed by the factor of the lattice
spacing, 
a reasonable guess for which might be 
around $10^{-16}$ (eV)$^{-1}$. 
It thus seems unlikely that this extra factor would 
affect calculations
at today's accelerator energies. 
It breaks Lorentz invariance explicitly. As 
mentioned before, 
this theory is only approximately Lorentz invariant.
Lorentz invariance is good at energies small compared to the inverse
lattice spacing.
From a more fundamental point of view, the rest 
frame of the crystal is a preferred frame and calculations should be
performed in that frame. However, observable effects of this frame
dependence are limited to very high energies. These are possibly
accessible through studies of cosmic rays (see sec. III).

To sum, building on the known equivalence 
of the Minkowski path integral to a 
stochastic process involving a complex Langevin equation, it 
has been shown that ordinary quantum field theory may result from the
dynamical critical behavior of an expanding phase boundary in a 
four-dimensional Euclidean space. In this picture, quantum 
fluctuations are actually thermal fluctuations in 
the higher dimensional space. 

\section{Special Relativity Realized Dynamically}

The underlying theory pictured above 
is a classical dynamical system lying
in the 4-D Euclidean space, governed by a universal Newtonian time.
It is proposed that Minkowski 
space is the result of restricting attention to the hypersurface
representing the phase boundary, and choosing local time to be the
spatial coordinate perpendicular to the moving boundary.
In calling it a Minkowski space, we are considering only a small
portion of the surface which can be taken to be approximately flat.
Globally, the spatial geometry is hyperspherical, and the space
is a positively curved pseudo-Riemannian space similar to the 
positive-curvature case of the Robertson-Walker metric of 
standard General-Relativity-based big-bang cosmology.

To show the emergence of Minkowski space locally,
a more detailed model is needed. 
If the phase boundary is considered the boundary between a liquid and
crystalline solid, with the solid growing into the liquid, then 
a reasonable model for the photon is the surface acoustic wave,
and for elementary fermions, screw dislocations in the crystal.
The surface acoustic wave is a propagating solution within the surface
that decays
exponentially away from the surface. 
It obeys a phonon-like 
dispersion relation, with a speed somewhat below that of shear bulk
waves. It is well known from the 1938 work of Frenkel and Kontorova 
\cite{frenkel} and of Frank and Eshelby in 1949 \cite{frank} that
screw dislocations obey the Lorentz contraction formula with the
speed of light replaced by the speed of transverse sound. In other
words, the pattern of crystal distortion that surrounds the 
dislocation becomes elliptical for a moving dislocation, with the 
strain pattern in the direction of motion shrinking according
to the Lorentz contraction formula. An ``object" made from an array
of such dislocations really does shrink in the direction of motion.
In addition, the effective mass of the dislocation grows with velocity
according to the relativistic formula (more precisely the energy
and crystal-momentum transform according to the 
Lorentz transformation)\cite{frank,weertman,nabarro}.
Therefore, screw dislocations are prohibited from being accelerated
beyond the velocity of transverse sound in a crystal, because the
kinetic energy becomes infinite in that limit. In a real crystal,
however, this limit can be exceeded by introducing a moving 
dislocation from an adjacent compatible medium where the sound speed is
higher. The supersonic dislocation rapidly decelerates to subsonic
velocities by emitting ``vacuum Cerenkov radiation"
\cite{weertman, nabarro}. It is also conceivable
to exceed the limit by violating the approximations of 
continuum linear
elasticity theory on which these results are based.
This relativistic behavior appears to be followed for any reasonable
dislocation model for which perturbations are subject to 
continuum linear
elasticity theory\cite{nabarro}. 
It is not immediately clear what the minimum
requirements are\cite{unziker}, 
but coupling to a single type of phonon with
a relativistic dispersion relation is necessary, 
and may be sufficient.
Coupling to other types of phonons is possible only if 
these either have the
same velocity or have an energy gap. The main point here is there can 
not be more than one limiting velocity for low-energy excitations. 
For instance, ordinary edge
dislocations obey a more complicated set of contraction equations
involving both the longitudinal and transverse sound 
velocities\cite{frank,weertman}.

Considering again the phase boundary universe model, if all matter is
made up of screw dislocations then the above considerations 
strongly suggest that measuring rods constructed from ``dislocation
arrays" will obey the Lorentz contraction. For now, consider only
observations made from the rest frame of the crystal. A measuring
rod will {\em physically} shrink if moving 
with respect to this frame along the rod's direction.
The Lorentz transformation also involves time, however. The Lorentz
contraction and mass increase certainly 
will have physical effects on clocks
that are constructed from moving dislocations. G\"{u}nther 
\cite{gunther} 
has investigated
using the breather solution of the sine-Gordon equation as a clock 
(sine-Gordon soliton kinks are a lower-dimensional dislocation model).
He finds such a clock slows with velocity in accordance with the usual
time-dilation formula.
If length and time standards are both based on solitons, full Lorentz
invariance ensues.

For our case, assuming only 
length contraction and observing from the crystal 
rest frame,
a simple light-clock where a flash
of light is given off and bounces off a mirror held by a rigid frame
to the light source, then back to a detector near the source, in either
transverse or longitudinal orientations, exhibits time dilation 
following the usual treatment in special relativity. However, although
the argument is the same, the assumptions are different. At this
point we have not assumed anything about moving frames of reference.
We are simply observing a moving rod and a moving clock from the 
rest frame of the crystal, where we know the speed of sound (light),
and know it is isotropic (we are always assuming an isotropic crystal).
This is all that is needed to demonstrate time dilation from Lorentz
contraction of the light-clock.
We notice that when observed from this frame, rods shrink,  and clocks
slow down due to physical, dynamical effects. Energy and 
crystal-momentum 
of dislocations also
obey relativistic equations\cite{frank,weertman,nabarro}. 

Now we ask what coordinate system is
a reasonable one for a moving observer to use? Of course, the moving
observer will use the shrunken rod to measure distance and the slow
clock to measure time  - what other reasonable choice does s/he have?
It is also natural for moving observers to choose their local
time coordinate to be along their own world line, and  spatial
hypersurfaces to consist of points all with the same time coordinate,
with synchronization performed using light signals.
The full forward Lorentz transformation, which consists not only of
scale changes inherent in Lorentz contraction and time dilation, but
also in the aforementioned axis rotations, ensues.
This now allows us to transform coordinates between the crystal rest
frame, and the natural frame of a moving observer. Inverting this
transformation is simply a matter of mathematics. As is well known
but seems to have been initially unappreciated by Lorentz,
this inverse
Lorentz transformation has the same form as the forward transformation,
with the relative frame velocity negated.  The point is that once 
the full forward Lorentz transformation is realized, fully reciprocal
special relativity results simply due to the mathematics of the Lorentz
transformation. 
In Einstein's special relativity, this is due to the symmetry of the
underlying Minkowski space - a kinematical symmetry. All frames are
exactly equivalent. Although our result is the same, conceptually it
is very different, since the Minkowski space has resulted from a 
dynamical symmetry of the moving boundary solution.
Unlike in the Einstein picture, the Lorentz contraction and
time dilation have different causes in different frames in this picture.
From the crystal rest frame, the shrinking of a moving rod and 
slowing of a clock are physical effects, caused by motion within the
stationary crystal. From the moving frame, the observation that a
rod and clock in the crystal rest frame {\em also}
appear to be shortened and
slowed are more of an illusion, created by using moving instruments,
and a bent reference frame, with its different notion of simultaneity. 
Because these points of view are conceptually different
(kinematic vs. dynamic symmetry),
Lorentz, Larmor, Langevin and others held on to the latter
view for some time after special relativity won 
acceptance\cite{miller}. 
In fact, the view of relativity given above is very similar to that
of Lorentz, who introduced the concept of local time given above.
The unobservability of the ether in this continuum theory eventually
led to the demise of this viewpoint. 
However, if the underlying medium is not a
continuum, but a lattice (which itself may lie in a continuum), then
at high enough energies differences between the stationary and moving
observer must eventually show up. This is because unlike the photon,
the phonon dispersion relation is not a straight line. 
For a linear isotropic material	in three dimensions
it is given by
\begin{equation}
\omega ^2 (k) = (2c/a)^2 (\sum _{i=1}^3 \sin^2 (k_i a/2)) .
\end{equation}
Surface phonons follow a similar dispersion relation. 
One has to get to within about 20\% of the maximum frequency
before the phonon curve differs from the photon curve by 
more than 1\%.
Above this point significant dispersion occurs. A very fast-moving
light clock which blue-shifted the light into this frequency 
region would 
show measurable deviations. The crystal rest frame will be the only
frame in which the speed of 
light at these high frequencies remains isotropic.
It therefore becomes an observable preferred frame - the ether 
is detectable.

These considerations suggest a number of ways that this theory could
be checked experimentally. Of course, the lattice spacing could 
always be made impossibly small, erasing all observable effects.
Observations of very-high-energy cosmic rays can put a lower bound
on the lattice spacing. Assuming the Plank relation $E=\hbar \omega$
(a possible origin of which is given in the next section) and setting
$\hbar = 1$, the definite identification of cosmic ray photons at
energies of a few times 
$10^{13}$ eV \cite{catanese}
means the high-energy cutoff of the photon
dispersion relation must lie above this, 
so probably $a < 10^{-14}$ (eV)$^{-1}$.
An interesting enigma in Cosmic Ray physics is the presence of
an ``ankle" in the spectrum  around $10^{18}$ eV, where the drop
in intensity with energy becomes less steep, along with the apparent
absence of the expected cutoff due to interactions with the 
cosmic microwave background radiation (CMB)\cite{cosmicray}. 
This Greisen-Zatsepin-Kuzmin (GZK)
cutoff \cite{gzk} is 
due to pion photo-production from interactions between
the cosmic ray particle (assumed a proton or light nucleus) and CMB
photons. This effectively limits cosmic rays of energy above 
$5 \times 10^{19}$ eV to a relatively short travel distance - within 
the local supercluster (photons and heavy nuclei are also limited by 
similar mechanisms involving starlight). 
However, the number of cosmic rays at this energy
and higher, although small in absolute event counts, does not show
any diminution from the earlier trend. 
In other words, there is no observational evidence for the GZK cutoff.
Another puzzle is that if
the very high energy cosmic rays do come from nearby sources, then,
they would be expected to point to within a few degrees of their 
sources, despite the deflection of magnetic fields, due to the high
momentum of the particles. However, there appears to 
be no correlation
with possible nearby sources.  
A photon energy cutoff in the range $10^{16}$ to $10^{19}$ eV
could invalidate the Lorentz transformation which is used to
derive the GZK cutoff from the known behavior in the center-of-mass 
frame\cite{ac}.
It would also affect the decays
of other high-energy particles, such as neutral pions. A $10^{20}$ eV
neutral pion could not decay into two photons, but at minimum into 
10,000 photons for a photon energy cutoff of $10^{16}$ eV. 
This would be highly suppressed due to the large 
power of the fine structure constant required. If the weak bosons
had similar cutoffs, then it seems the neutral 
pion could be made almost
stable above a certain energy.
Decay of the neutron could be similarly suppressed. If one or more
of these neutral particles could travel cosmological distances above
a certain energy threshold, it could possibly explain the ankle,
due to the addition of a new species to the particle mix.
High-energy neutral particles should point toward their sources  
even at great distances, since they are not much affected by magnetic
fields (there is still some effect through magnetic moments).
Interestingly, Farrar and Biermann have shown 
that the observed directions of some
of the highest energy events can be correlated with distant 
quasars\cite{farrar}.
This would be consistent with the scenario sugested here.

\section{Photons as Surface Acoustic Waves}

Generically, surface acoustic waves (SAW's) 
traveling in the $x$-direction
on a solid surface at $z=0$, with the solid occupying the half-space
$z<0$ takes the form\cite{rayleigh,saw}
\begin{equation}
u_j = u_{0j} e^{i(kx-\omega t)+\kappa z}
\end{equation}
with $k$, $\omega$, and $\kappa$ real, and $\kappa \propto k$ 
at least for small $k$. Here $u_j$ is the elastic displacement field
for the solid, which is defined only for $z \leq 0$.
Typically most of the energy in surface waves is confined to a region
within a few wavelengths of the surface. 
The most common SAW is the Rayleigh wave, first described by 
Lord Rayleigh in 1885\cite{rayleigh,saw}.
It is polarized in the saggital plane 
(perpendicular to the surface), and consists of motion that is both
transverse and longitudinal. It is dispersionless in the continuum
version and has a typical phonon dispersion law on the 
lattice\cite{latdyn}.
This wave does not seem to be a promising one to model the photon
after, however, since it has only one polarization, regardless of the
dimensionality of the surface. The Rayleigh wave is the only 
type of surface
wave for the simplest case of a flat linear elastic half-space.
However, if the surface is allowed to have properties different
from the bulk, such as a different density, elastic constant, surface
tension, curvature, roughness, piezoelectricity, magnetoelasticity,
etc. then another surface wave 
will usually exist, a Love wave\cite{viktorov,maugin,murdoch,farnell}.
This wave, originally derived
for a finite slab of different material deposited on the 
half-space\cite{love},
has a shear-horizontal (SH) polarization, 
thus for a three dimensional
surface would have two transverse polarizations. The Love wave
also exists for a thin surface layer such as a thermodynamic
phase boundary\cite{maugin, tiersten, kosevich}. 
It can be seen as a perturbation of the SH surface
skimming bulk wave (SSBW) that 
exists even for the simple half-space. The SSBW
is a wave that does not decay below the surface. This solution is
unstable with respect to virtually any
surface property that retards the wave speed near the surface, 
which will
turn it into an SH surface wave with exponential decay away from 
the surface, i.e. a Love wave\cite{viktorov, maugin}. The Love wave is 
somewhat dispersive, due to 
the introduction of a quantity with dimensions of length that
characterizes the surface-layer thickness. 
However if this is
no more than a few lattice spacings, then the dispersion is 
similar to that of an ordinary phonon. 
Adding a liquid to the external space complicates but does
not significantly change the situation.
However, the case being envisioned here has a more 
complicated boundary 
condition than has been considered in the surface-wave literature,
since the boundary is growing, perhaps rapidly. This can perhaps
be treated by the method of virtual 
power\cite{vpower}, and is briefly
considered by Maugin\cite{maugin} and also by Kosevich and Tutov
\cite{kosevich}. 
The transfer of momentum to the ``new surface"
of the growing crystal will modify the usual traction-free boundary 
condition of the Raleigh-wave solution. It is this latter 
boundary condition 
which prevents the SH polarization from existing 
in the simple half-space\cite{saw}. 
The violation of this boundary condition
by the growing crystal is further evidence that SH waves probably
do exist in this case.

Since the weak interactions also need to be accounted for, 
probably more structure needs to be incorporated into the model.
If we imagine the elementary molecules to be non-spherical, then
they have their own non-trivial symmetry group, compatible with but
distinct from that of the crystal. This basis symmetry group could 
account for internal symmetries. For instance, if the molecule can 
be represented by a 4-d vector with nearest-neighbor Heisenberg-like
interactions, then an SO(4) symmetry (isomorphic to 
SU(2) $\times$ SU(2)), which may be partially broken by
other interactions, will exist. The spontaneous 
breaking of this symmetry
will result in surface magnons\cite{cottham}. 
These come in both acoustic and optical
varieties. The surface magnons can also mix with surface elastic
waves through the magneto-elastic effect, reminiscent of electroweak
unification.
These possibilities need to be examined in detail - they are mentioned
here to indicate the rich possibilities for model building that occur
in surface modes. 
It is also worth noting that a 
promising approach to incorporating chiral fermions
on the lattice, necessary for a lattice approach to the 
weak interactions,
incorporates a fourth spatial dimension, with the chiral fermions
living on a domain wall\cite{rubakov, chiral}. 
The picture of the universe presented here
seems ideal for the realization of this mechanism.
 
A universal property of all surface modes is the 
exponential decay
as the bulk is entered, characterized by a decay length 
proportional to 
the wavelength. This property can be used to derive the Plank relation
$E=\hbar \omega$, perhaps the most fundamental equation of quantum
mechanics, from the equipartition theorem. 
If we assume that all elementary degrees of freedom are
thermally excited (actually not a completely 
good assumption due to conservation
laws and partial non-ergodicity - see discussion below), 
then the equipartition theorem will give 
equal energy to each harmonic degree of freedom of 
amount ${\rm k}T_4$,
where k is Boltzmann's constant and $T_4$ is the 4-d temperature.
For a surface wave with decay length $\kappa = bk$, where $b$ is 
a constant, taking into account the energy of a wave being 
proportional to its square, one has an energy 
depth profile (1-d energy density)
\begin{equation}
E = E_0 e^{2\kappa x_4}
\end{equation}
where $x_4$ is taken to be zero at the surface, and becomes negative
inside the medium.
The energy in the monatomic surface layer itself is given by $E_0 a$.
The total energy can be computed by 
\begin{equation}
E_{\rm tot} = \int_{- \infty}^{0} E dx_4 = E_0 /(2\kappa)
\end{equation}
Setting $\kappa = bk$ (proportionality of decay length to wavelength),
$\omega =ck$,
and the total energy to ${\rm k} T_4$, one can solve for the surface
layer energy, $E_0 a$, now denoted $E_3$
\begin{equation}
E_3 = (2ba{\rm k}T_4/c)\omega
\end{equation}
which is the Plank relation if $\hbar = 2ba{\rm k}T_4/c$. This is 
consistent with the thermal explanation of quantum 
mechanics given above,
namely that $\hbar$ is essentially the 4-d temperature, 
with the necessary
factors of $a$ and $c$ to fix the dimensions.
The essential feature which gives higher frequency excitations higher
energy on the surface is the higher concentration of SAW 
energy near the
surface, compared to lower frequency excitations which are more 
spread out in the fourth dimension. Thus equal sharing of energy in
four dimensions naturally leads to unequal energies on the 3-d 
surface, as embodied in the Plank relation.

Not all surface modes will necessarily become excited for two reasons.
First is the effect of global conservation laws, and second is the 
probable lack of full thermodynamic equilibrium. Consider the liquid
degrees of freedom directly above the growing surface. These are
presumably in thermal equilibrium in their liquid environment. When the
surface arrives, they are rather suddenly thrust into a new
environment with modified interactions due to 
the translational symmetry
breaking of the crystallization. They therefore do not have much time
to adjust to these new conditions by the time they 
can be considered part
of the new crystal surface. Eventually they reach a new equilibrium
state well after the surface has passed and they become 
part of the bulk.
Thus the surface degrees of freedom are in a transitional state.
With the arrival of crystalline order comes a new conserved quantity,
the crystal momentum. It is a consequence of the remaining discrete
translational invariance but technically is a permutation invariance
of the atomic position variables, resulting in conservation of 
wave number
for phonons\cite{ash}. 
Since this is only a single global constraint (or three
constraints in three dimensions) it would not appear to 
limit the allowed
random excitations much. However, satisfying global 
conservation laws
requires global correlations, and these take a long time to establish.
Therefore one expects to have to satisfy conservation laws locally.
This means that one should not expect widely separated
thermal excited phonons whose wave vectors happen to add to zero. 
Rather one expects standing waves or standing wave packets where
the zero net wave vector requirement is met pairwise and locally.
Thus the random vibrational 
thermal energy of the liquid will re-organize on the
surface primarily as such standing waves, with the amplitudes of 
constituent travelling waves locked, and phases randomly fluctuating.
This may represent the zero-point energy of the photon field.
Since each travelling wave mode is not independently excited, the
net energy assigned to each is one-half that of the standing wave,
i.e $\frac{1}{2}\hbar \omega$. 
 
However, if a travelling propagating wave already exists on the surface
(perhaps excited by dislocation interactions etc.) then it will
be preserved by the crystal momentum law, and, since it is now 
an allowed excitation can exist independently of the standing wave
and be given the full equipartition energy of $\hbar \omega$,
on top of what it gets from the zero-point excitation.
One wonders how multiple photon excitations can arise in such a 
picture, which will lead to a discussion of resonances in coupled
oscillator systems. Before embarking on that, it is worth noting 
that all of the discussion here concerning zero-point energies and
photon excitations is in one sense unnecessary, 
since once the formal
equivalence of the stochastic evolution of the phase boundary and
the quantum system is accepted, one can merely plug in the QED or
standard model Lagrangian and obtain the equivalent 
Langevin equation for the
phase boundary
evolution, which will, 
due to the above equivalence, 
necessarily include all of the known particle
excitations and quantum effects. 
The discussion here is therefore not to prove that 
each feature of quantum mechanics is included, 
but rather to illustrate
how each quantum feature might be 
manifested in the phase boundary evolution.
In a similar sense, energy quantization is not as apparent in the 
path integral formulation of quantum mechanics as it is in the 
canonical formulation, but it has to be there, and can be seen from
multiple poles of the propagator.

A collection of coupled oscillators, even if somewhat non-linear, is
generally not ergodic. This was demonstrated by the famous computer
simulation of Fermi, Pasta and Ulam in which they coupled 64 harmonic
oscillators with non-linear couplings, expecting to see the approach
to equilibrium\cite{fermi,rasetti,robertson}. 
Instead they found that most modes remained unexcited
with energy pouring back and forth between a few modes as in the 
Wilberforce pendulum - in other words a limit cycle as opposed to
chaos. The only modes that participated were those that met or were
very close to the
resonance condition
\begin{equation}
\sum_i n_i \omega _i = 0
\end{equation}
where the $n_i$ are integers\cite{rasetti,robertson}. 
This was later understood in terms of 
the KAM theorem (Kolmogorov, Arnol'd, Moser), 
which essentially states that for
small non-linearities only regions near resonant surfaces in phase
space will get occupied. Full chaos only ensues when these resonant
regions grow large enough to be overlapping\cite{ford}.
Phenomena such as down-conversion or harmonic
generation in the presence of small non-linearities
can be understood in terms of the resonance condition. The n-photon
state takes the form of an 
n$^{{\rm th}}$ order resonance from this point of view. 
The integers in
the resonance condition are the correspondents to energy quantization.
The degree of excitation is consistent with that of a single 
n$^{{\rm th}}$ harmonic photon with which the state is resonantly
linked.
Therefore it seems plausible that the Plank relation and full photon
spectrum, including zero point energy, does have a realization
in the propagating modes of the phase boundary as it moves through
the random medium.

\section{Quantum Superposition and Measurement - Zitterbwegung 
and Spontaneous Symmetry Breaking}
The picture described above treats quantum fluctuations as
thermal fluctuations in the 4+1 dimensional space.
In such a picture quantum tunneling is explained
classically as thermal activation, i.e. due to a random 
kick of extra energy which results
from thermal contact with the liquid and solid phases.  Due to such
thermal fluctuations, energy is not conserved over short time periods;
it is conserved only in the average over time.
Thermal fluctuations may create a kind
of zitterbewegung -- very rapid variation at small scales, that
enforces the uncertainty principle and allows for
superpositions. The ensemble average implied in a
quantum expectation value is replaced by a time average. 
For rapid fluctuations which cover the ergodic subspace in times
short compared to the time between measurements, these should
yield identical results. Over very short time periods, additional
correlations may appear in the time-averaged case, since the 
classical system is in a particular state at any one time, so
subsequent states will retain some memory of previous states.

This more classical evolution
affords the opportunity to
explain the quantum measurement process 
as a spontaneous symmetry breaking event.
Anderson has suggested that measuring devices incorporate
spontaneous symmetry breaking in their operation\cite{anderson}. 
Ne'eman
has also espoused this viewpoint. In addition, he has shown that
EPR type correlations can occur in classical systems with
gauge symmetries, with the gauge connection enforcing long-distance
correlations among fluctuating variables\cite{neeman}.
More detailed models have been considered in \cite{zim} 
and \cite{ssb}.

When a classical statistical mechanical system undergoes a spontaneous
symmetry breaking, the ergodic phase space 
splits into non-communicating
subspaces. From that point on, the system remains trapped in one of
the subspaces. Which subspace is chosen is simply determined by
the subspace the system happened to be in 
at the time of symmetry breaking.
A measuring device is postulated to be any device 
that can couple its order parameter to a quantum system
and that includes a control
that can initiate spontaneous symmetry breaking of 
that order parameter. The measuring device, 
originally with an unbroken symmetry, couples
to the system under study becoming strongly 
correlated with it. Then
an adjustment is made to the potential of the 
measuring device which initiates spontaneous symmetry 
breaking. 
The measurement takes place at this time,
when the ensemble
of possible future states of the combined system splits
into non-ergodic subensembles corresponding to the
possible values of the order parameter,	also corresponding
to possible values of the measured quantity.
Future evolution is confined to a single subensemble
in the usual manner of a classical symmetry-breaking phase transition.
In this picture measurements are well defined, the 
collapse is a physical event, and a clear distinction exists
between what constitutes a measuring device and what does not.
The symmetry breaking barrier does not even have to be infinitely
high - all that is required is that the tunneling time of the
post measurement state to be long
compared to the time scale of the experiment.
This is in contrast to what occurs if the same concept of spontaneous
symmetry breaking is applied 
to explain measurement
in standard quantum mechanics. Here
it is difficult to see how even spontaneous symmetry breaking can
break the superposition, especially if measuring devices are finite
so tunneling probabilities are not quite zero (e.g. ref. \cite{zim}
still uses the Everett interpretation to deal with the ``collapse").
Nevertheless it is
assumed that when the universe undergoes a cosmological phase 
transition
it does not end up in a superposition of the possible outcomes but
rather ``measures itself" so as to fall into a single vacuum.
In the new picture given here, an event like this is in the same
category as a measurement and the result of both 
is a physical collapse of
the available future phase space. 

Because the ``current past'' is continuously undergoing 4-d thermal
fluctuations, it is only frozen to the extent that the ensemble is
limited due to spontaneous symmetry breaking. Thus questions such as
``which slit did the electron go through'' or ``which direction was
the spin pointing'' are as meaningless here as they are in standard
quantum mechanics. This is because the details of history 
are continuously
being rewritten as both the current past and present fluctuate. 
Only to
the extent that the ensemble is limited by spontaneous 
symmetry breaking
can one make definite statements about past events.
EPR (Einstein-Podolsky-Rosen) states, which consist of two separated
spins in a net spin-0 state, can only undergo correlated fluctuations
which obey the global angular-momentum conservation law. 
The spin direction of each particle will fluctuate in such a way that
its partner fluctuates oppositely. Measurement of either spin is
performed by spontaneously breaking the spin direction symmetry,
after which a barrier will exist preventing further fluctuations of
either particle. Such a process was envisioned in \cite{neeman}.
Such non-local correlations may seem odd, but
they are formed by the causal process of separating the particles,
and do not violate causality (causality is arbitrated from the crystal
rest frame, where universal and local time are equivalent).

\section{Dislocations as Candidates for Elementary Fermions}
Dislocations, particularly screw dislocations and their variants,
provide a rich building ground for models of elementary particles.
In this section a detailed model will not be attempted, but rather
the general problem of extending the screw dislocation into four
dimensions will be discussed, which will result in a 
four-dimensional string.

The idea of representing elementary particles as dislocations in 
a medium is a rather old one. Burton talked of ``strain-figures"
that could move through 
a medium and interact\cite{burton}. Although most 
19th century physicists considered matter to be separate from
the ether, Larmor suggested the possibility of matter particles
being singularities in the ether itself and sought a unified theory
of matter and radiation through the properties of a single 
medium\cite{larmor,ether}.
In more recent times the modeling of elementary particles as
topological solitons (a type of dislocation) has intrigued many,
with the Skyrmion picture of the nucleon being perhaps the
most successful.

The screw dislocation in three dimensions 
has a number of features
that liken it to an elementary fermion. The left and right 
handed versions
can be pair-produced or annihilated, and their elastic interactions
have a number of electromagnetic analogies,
the most often cited being to magnetostatics\cite{nabarro,
eshelby,dislocations,dewitt}. 
Although double dislocations
are not totally prohibited, they are very unfavorable energetically.
Screw dislocations
are, of course, line defects, so cannot be compared directly to
point particles. One is tempted to 
interpret the line defect as a world-line.
However, this implies an extension to four dimensions. An isolated
screw dislocation is unfortunately not an option in four dimensions.
This can be seen in a number of ways. If one circles a 
screw dislocation
in three dimensions, then one finds after a single loop that one has
advanced one lattice spacing to the next sheet of atoms 
in the direction
of the dislocation. The degree of non-closure of the 
loop is represented
by the Burgers vector of the dislocation, which for a screw 
dislocation on
a cubic lattice is one lattice spacing long and in the same direction 
as the dislocation, or opposite for an oppositely-handed dislocation.
Although the transition is gradual, the point on the loop at which one
can be deemed to be on the next level can be arbitrarily defined - the
set of these points for all possible loops 
is called the Volterra surface. The freedom of choice
of the Volterra surface can be thought of as a form of 
gauge invariance.
There are many atoms far from the dislocation which have moved
some fraction of a lattice spacing from their original 
lattice positions,
but there is not much stress associated with this since 
all of the neighboring
atoms have moved a similar amount. Stresses are concentrated only
around the dislocation line. If one tries to embed this structure into
a non-dislocated 4-d lattice, then those atoms far from the dislocation
which are shifted from their original lattice positions by near 1/2 of
a lattice spacing will fit badly the undislocated lattices 
adjacent to them in the fourth dimension, where the atoms are all at 
their original undislocated positions. The energy of such a structure 
is proportional to the four-volume - it is no longer a one-dimensional
dislocation. The other way one can see there is something wrong
in simply promoting the screw dislocation to four dimensions, is that 
a loop apparently surrounding the dislocation can be moved into the 
fourth dimension, where the dislocation does not exist,
and shrunk to a point. Thus there is no longer a consistent topological 
classification of this object.

The screw dislocation can be extended into the fourth dimension by 
copying it onto each successive 3-lattice as the fourth coordinate is
changed. This produces a wall of identical dislocations. The solution
is translational invariant in the fourth dimension and involves no
new stresses, since each atom is exactly one lattice spacing away
from its neighbor in the fourth direction. However, we now have a 
domain wall in 4-d or line in each 3-d slice. 
For an elementary particle we want
something closer to a point in 3-d. An obvious solution would be to
wrap the domain wall around onto itself into a small tube, the 3-d
cross-section of which would be a string. The bending of the wall
introduces stresses which favor a larger string, but this is opposed to
the ordinary screw stress proportional to the string length, so there
is the possibility of a stable equilibrium size. Going back to the 3-d
cross-section, the Volterra surface is any surface 
bounded by the string.
A loop that threads the string will pass through the Volterra
surface and detect the dislocation. 

Assuming a planar loop in the 3-d cross-section introduces a direction,
the spatial normal to this plane (the temporal direction 
is also normal).
This suggests the possible interpretation of a spin direction. 
Another strong possibility is that the constituent screw dislocations
are not straight but form spiral helices. Ordinary screw dislocations
often take helical form through a process that involves absorption
of interstitials or vacancies\cite{nabarro,dislocations,dewitt}. 
The plane of the helix introduces another
spatial direction which could be related to spin or spin precession.
Interstitials are important in that they introduce curvature into the
crystal\cite{kroner}. 
Such curvature is absolutely necessary to produce the
large-scale hyper-spherical spatial geometry inherent in 
the cosmological
scenario outlined above. It also allows a connection between particle
properties and gravitation.

One additional property of crystal dislocations that may provide an
intriguing parallel to the strong interactions is 
the existence of partial
dislocations\cite{nabarro, dislocations}. 
Under favorable conditions, a dislocation may split
into two or more  partial dislocations with fractional Burgers vectors.
Such objects cannot exist in isolation since they would involve
dislocating the entire lattice - resulting in infinite energy. 
These partial dislocations are linked by a sheet containing a stacking
fault, which produces an attractive force proportional to the sheet
area (the partial dislocations also repel each other through other
elastic forces, resulting in an equilibrium separation). The possible
analogy between quarks and partial dislocations, with gluons being
related to the associated stacking faults is compelling. Confinement 
and fractional charge are
inherent and linked properties 
of these configurations. Another common feature
of dislocations in ordinary crystals is the formation of dislocation
networks. These are ordered or disordered collections of either partial
or full dislocations and anti-dislocations, 
with zero net Burgers vector.
New kinds of dislocations can be defined from defects in an otherwise
ordered dislocation network. 
For instance the chiral condensate could be modeled as a network of
partial dislocations and associated stacking faults. Nucleons and
mesons could then be modeled as dislocations and excitations of this 
underlying network, which is reminiscent of the Skyrmion approach.
Ordered dislocation networks can  have dislocations 
which can form
an ordered network which can itself have higher-order dislocations,
producing a possible hierarchy of dislocations several levels deep.

\section{Gravity as Elasticity of Space}
The similarities between 
the General Theory of Relativity and the theory
of elasticity have been remarked upon by many authors. Sakharov
spoke of relating General 
Relativity to a ``metrical elasticity of space"\cite{sakharov}.
Kokarev has likened space-time to a ``strongly-bent plate"
\cite{kokarev}.
Several authors have developed three-dimensional continuum models of
dislocations that resemble three-dimensional gravity\cite{3dg}. 

Screw dislocations themselves do not result in curvature - rather they
introduce torsion into the lattice, since an observer circling a screw
dislocation finds themselves transported forward, along the dislocation
direction. Two sources of curvature have been 
put forward - disclinations
and extra matter (primarily interstitials). Disclinations are 
large angular defects. For example, the pentagons in a geodesic
dome can be thought of as disclinations in an otherwise flat
hexagonal tiling, and produces obvious curvature in the surface. 
The problem with
disclinations is that they 
produce curvature only in large finite chunks,
rather than building up from many small pieces. 
So, whereas a disclination
is a good model for a cosmic string\cite{vilenkin}, 
it is not a good candidate for
an elementary particle. We are therefore left with the extra matter
concept, which has been championed by Kroner\cite{kroner}. 
In Kroner's theory,
the geometry of the resulting 
continuum model is characterized by both curvature,
the source of which is extra matter, and torsion,
which is caused by dislocations\cite{dewitt,kroner,extram}.
The obvious four-dimensional generalization would be the 
Einstein-Cartan-Sciama-Kibble theory of gravity, which 
supplements the
usual Einstein equations with an equation relating spin 
density
to the torsion tensor\cite{hehl,gokeler}. 
Torsion effects are too small
to be detected experimentally, so this theory is, so far, 
experimentally
indistinguishable from General Relativity.
In order to satisfy the equivalence principle, the absorption
of interstitials by dislocations mentioned above would have to be
a universal property, with the degree of absorption proportional
to the energy, so that curvature could couple to the energy-momentum
tensor. It is not clear that this would necessarily happen,
however it could be forced by symmetries, since due to the Bianchi
identity the Einstein tensor can only couple to a conserved quantity.
Not all interstitials are necessarily absorbed. Unabsorbed 
interstitials are an intriguing dark-matter candidate. Unlike 
ordinary particles they do not persist, since they are true 4-d
point defects. Their behavior is more like that of 
instantons. Their fleeting
existence could make their detection difficult other than through
their gravitational effects.

Regardless of the details of the gravitational theory that results,
it will necessarily be a quantum theory of gravitation. This is 
because the evolution of the spatial hypersurface is influenced
by the thermal fluctuations in the surrounding medium which are the
source of quantum fluctuations in this picture. One certainly expects
thermally induced curvature fluctuations. However, if the elementary
lattice spacing is much larger than the Plank length, it is likely
that such curvature fluctuations would be small, and gravitation would
remain, in practical terms, a largely classical theory. As distances
approached the lattice spacing, then the continuum theory 
(presumably a generalization of General Relativity) would have to be
replaced with an appropriate lattice theory, 
just as continuum elasticity
theory can be used for a crystal only for distances large compared 
to the lattice spacing. Of course, just as in 
ordinary crystallography, the lattice
theory itself may be based on an underlying continuous space.
The resolution of singularity problems 
in general relativity are more likely to come from the transition
to an appropriate lattice theory than from the incorporation of
quantum effects, unless the lattice spacing is of order the Plank
length or smaller.

\section{Cosmological Consequences}
The model of an expanding phase boundary provides good explanations
for some cosmological puzzles but introduces additional problems
as well. Phase nucleation is a common way for structure to arise
from chaos spontaneously. It naturally creates an expanding universe
starting from a very small but not infinitesimal seed. Only if
the initial fluctuation is above a certain minimum size, will the 
crystal grow - otherwise surface tension effects will remelt it 
back into the liquid. There would not seem to be a horizon problem
because there is plenty of time before the big bang to establish
causal contact, thermodynamic equilibrium etc. Also there is 
the ``flatness
problem" which, in a non-inflationary universe, requires a careful
fine-tuning of parameters to create a universe as long lasting as ours
which nevertheless has a reasonable matter density and is close
to being spatially flat in the present era. Phase transitions only
occur when there is a fine tuning between various terms
in the Hamiltonian, so a system undergoing a phase transition is
already {\em naturally fine tuned} 
between forces that favor the transition
and those that don't. The other ingredient this model likely has
which could reduce the need for fine-tuning would be dissipation,
which could tame runaway solutions like inflation.
In general, surface growth which is not diffusion-limited is 
controlled by the volume energy (which results in the liberation
of latent heat), surface tension, and dissipation. 
The outward pressure from the volume energy
takes the form of a repulsive cosmological constant and
the 3-d surface tension may act like the ordinary spatial
curvature term, but it is not 
immediately clear
how to take dissipation into account within the standard
Friedmann models. Comparison to ordinary phase transitions 
would suggest a period of slow growth at first, which accelerated
as the surface term became less important, finally approaching
a steady state constant growth rate.
One can also consider the possibility that the background conditions
responsible for the supercooling could vary over time. If this is
allowed then a more complex growth-rate history could be accommodated.

An intriguing possibility for matter generation would be collisions
between different crystal universes. Where crystals join, a lot
of dislocations are formed. The join-boundary of two 3-d surfaces
is a 2-d surface. Therefore, dislocations produced in such collisions
would be distributed on 2-d surfaces within the combined 3-d surface
of the joined crystals.
Interestingly, matter in the universe is primarily distributed on
a network of 2-d surfaces surrounding large voids. One can imagine 
this resulting from the twisting and folding of the join-boundary
of a single cosmic collision or from a number of such events. 

This scenario may have difficulty explaining both the uniformity of
element abundances and of the cosmic background radiation. Helium 
could be produced in the cosmic collisions referred to above  
in much the same way as in the hot big-bang, but conditions would
likely vary somewhat from place to place. A single large collision
might be able to produce a fairly uniform result. 
Cosmic collisions, in 
addition to creating matter in the form of dislocations would also
produce a lot of thermal radiation. Again, this could be fairly 
uniform for the case of a single large collision. This scenario
shares some features with the colliding-branes string-model picture
recently proposed by Khoury et. al.\cite{ekp}, although the geometry
is rather different.

\section{Discussion}

At first glance, the idea that space could be crystalline would seem 
at odds with the notion of spatial isotropy. Wouldn't the axis
directions create preferred directions in space? For distances large
compared to the lattice spacing, this is not necessarily so. 
For instance, the long distance behavior of an isotropic crystal
(one with isotropic elastic constants) is well approximated by 
isotropic linear elasticity theory which has full 
rotational invariance.
Another example is lattice gauge theory. Here forces along axis
directions differ from those along 
non-axis directions at short distances,
but full rotational symmetry emerges at distances large 
compared to the
lattice spacing. The longer lattice paths in diagonal directions
are exactly compensated by the larger multiplicity of such paths.
Also the surface of a growing crystal is more labile 
than the interior, resulting in features that are less ``solid".
For instance, even sessile dislocations can move through growth,
via formation of kinks and jogs, though they are essentially 
locked in place once formed. Glissile dislocations (those that can
move freely through the crystal) may themselves essentially stop
in the bulk by transferring all of their 
momentum to the ``growth tip"
through a mechanism similar to a Newton's cradle onto which balls
are added continuously, or a whip with a growing tip.

The similarities between condensed matter physics and particle physics
are many. \linebreak Phonons are surprisingly similar to photons. They can be
thought of as Goldstone bosons resulting from the breaking 
of translation
invariance, or as gauge fields relating to the remaining discrete
translational invariance, which due to lattice periodicity, may be
represented by an angular order parameter\cite{anderson}. 
The counterpart to the 
Higgs mechanism is the plasma mechanism\cite{anderson2}. 
Even the chiral properties
of the weak interaction may have an analog in the behavior of 
$^3$He-A\cite{volovik}. 
Several gauge theories of dislocations have been 
proposed\cite{edelen,kleinert}.
What is being proposed here can be thought of as going all the way
with this program, namely hypothesizing that particle 
physics {\em is} condensed matter physics. The main 
experimental signature of such a proposal, regardless of the details,
would be the effects of a finite lattice spacing. 
Besides the dramatic cutoff of gauge boson spectra above a certain
energy, one can look for effects of dispersion near the cutoff.
The lattice also makes all ultraviolet divergences finite, which
will introduce small effects in higher order corrections. 
This also adds impetus to proposals that a serious effort be made
to search experimentally 
for violations of Lorentz
invariance\cite{lorentzviol}. The effects of living on a 
physical lattice are somewhat different from string-inspired
Lorentz-invariance violations.
The other experimental signature this scenario has
in common with other extra-dimension scenarios is the possibility
of energy conservation violation beyond the statistical
violation already discussed and interpreted
as quantum fluctuations\cite{rubakov}. 
One can imagine the possibility
of a high-energy interaction 
radiating a longitudinal phonon into the bulk, for instance,
which would 
look like a missing-energy event. This can be made rare by
either a very weak coupling to these modes or by giving them a
low frequency cutoff (a mass). Radiation forward into the liquid could 
be prevented by having the surface growing at a rate exceeding 
the sound speed in the liquid. 
Another possible experimental signature to look for would be effects
of dissipation including lack of time reversal invariance. Although
conservation laws may prevent dissipation on the surface itself,
the bulk phases undoubtedly are dissipative. The moving phase
boundary breaks time reversal invariance spontaneously. Both T and
CPT invariance could be broken.

A final note concerning time in this theory is the special role
played by the present. The edifice of the universe is constructed
at the present surface from material provided by the undifferentiated
current-future (liquid) state. 
The past, being a solid, is more fixed, though still can undergo
some fluctuations.
Present, past and future are different, distinguishable phases.
This would seem to conform 
with our personal
experience better than the picture presented in special relativity,
where the present is not distinguished, and the future seems as
well-formed as the past. 
Indeed, according to Einstein, 
``For us believing physicists, the distinction between past, present,
and future is only an illusion, even if a 
stubborn one\cite{hoffman}."
Davies states, ``The four-dimensional space-time of physics
makes no provision whatever for either a `present-moment' or 
a `movement' of time\cite{davies}."
Quantum mechanics could play a possible role in blurring the future
in the standard picture, 
but this depends on a definite resolution to the measurement problem.
The phase boundary scenario, in contrast,
matches well with the ``process philosophy"
concept of time
as advanced by Whitehead, who talks of a ``concresence" unfolding at
the present where the indefinite future is molded into a definite
past\cite{whitehead}.

The preferred frame offered by the crystal rest frame also gives
a different point of view for causality arguments. One can imagine
the possibility of interactions that occur by faster-than-light 
mechanisms, just as a bullet can exceed the speed of sound in an
ordinary crystal. Although in some frames of reference, cause may
appear to precede effect, this will never occur in the crystal
rest frame, regardless of interaction speed. Since Lorentz invariance
is only approximate, all frames are not equivalent. The 
correct result is that observed in the preferred frame. Thus there is
no longer a paradox created by faster-than-light interactions by which
one could travel backwards 
in time and kill one's grandfather, for instance.
Time always goes forward and effect follows cause
in the crystal rest frame.
Of course there
is no evidence that any interaction or particle can exceed the speed
of light, and ordinary dislocations probably can not, as previously
discussed, but the removal
of this causality paradox opens the door to such a possibility a crack
wider.

\section{Conclusion}

At first glance this theory appears to be an anachronism - a
neo-Lorentzian classical ether theory. Modifying special relativity 
and reintroducing an ether are probably the last thing that 
would enter the
mind of a 20$^{\rm th}$ or 21$^{\rm st}$ century physicist, 
followed perhaps by a classical
explanation of quantum mechanics. However, there are many ways 
in which
this theory fits with modern ideas. The idea that we live on a 
membrane is
becoming popular in string theory, and also for introducing chiral
fermions into lattice gauge theory. Stochastic quantization, 
though never
fully accepted in the realm of quantum mechanics, came very close to
giving a classical statistical-mechanical explanation of quantum
fluctuations. The very many analogies between particle physics and
condensed-matter physics, 
especially in the realm of gauge field theories
and spontaneous symmetry breaking, has 
led to tremendous sharing of ideas
from one field to the other. The main difference between these 
is simply
between relativistic and non-relativistic spacetime symmetries. 
One other difference is that, whereas in elementary particle physics
gauge symmetries and Goldstone bosons are usually considered to be 
essentially separate mechanisms for producing massless particles
(which conflict in the Higgs mechanism to give a mass), in condensed
matter physics the gauge particles (phonons) can themselves be 
pictured as a type of Goldstone boson associated with the breaking
of translational symmetry. A vector order parameter yields a vector
Goldstone boson and a tensor order parameter (associated with breaking
of rotational invariance) should produce a tensor Goldstone boson
(the graviton). Gauge invariance is actually born from the ambiguities
in defining the unperturbed lattice\cite{anderson}. This economy of
ideas (gauge bosons as Goldstone bosons) is appealing and may help
to explain why some symmetries are gauged and others are not. To
benefit from this analogy, however, an 
ether-type background would appear to be
necessary to provide the required translational symmetry breaking.

The idea
of a moving, expanding, phase boundary, where relativistic space-time\linebreak 
emerges as a dynamical symmetry, integrates these ideas into a
coherent picture; the big-bang and gravity, from the flexible 
geometry of the interface, are incorporated almost for
free. The quantum measurement process is also vastly clarified in this
picture as a consequence of spontaneous symmetry breaking.
Ideas from chaos and ergodic theory, 
non-equilibrium statistical
mechanics and dynamical critical phenomena play an important role. 
When these are added, a classical theory 
doesn't look so classical anymore. 

    Not much has been mentioned in this paper concerning 
the base-theory. What new sets of 
even more elementary particles (called elementary atoms above) and
forces must be postulated for the underlying base-theory,
upon which the
dislocations and surface waves (our
current set of elementary particles in this picture) can be
built? Hopefully it is a simpler set than we currently have in the
standard model. It seems 
possible that one or two types of elementary
atoms, combined with a {\em short-range} force, 
repulsive at small distances and
attractive at long, could be enough. Some model building seems 
to be in order, 
starting with extending simple crystal models to four dimensions. 

   In closing, 
one cannot help but speculate whether Einstein would have
liked this idea. Certainly he may have disagreed with the
reintroduction of the ether which 
he so strongly fought against, as well
as the notion of a preferred frame. 
However from the point of view of the
more fundamental base-theory there is no preferred frame --  it is
introduced through a spontaneous symmetry breaking, so the
principle of relativity is safe there. Einstein's
discomfort with the inherently probabilistic nature
of quantum mechanics is well known, 
so perhaps he would be
pleased with the 
application of his ideas on Brownian motion toward the
explanation of quantum fluctuations, 
as well as the replacement of quantum
mechanics with a 
deterministic (though chaotic) theory. Finally, there is
the essentially geometric basis for electromagnetism and 
possibly all
interactions through the picture of a dislocated crystal. 
This bears a
rather strong resemblance to unified field theories that 
he worked on in
his later years. On balance, this theory 
seems relatively in concert with
the ideas of Einstein.


\begin{thebibliography}{99}

\bibitem{grady98} These ideas were introduced in M. Grady,
gr-qc/9805076; see also M. Chown, \textit{New Scientist}, {\bf 161}, 42, 1999.
\bibitem{rubakov} The idea of the universe as a domain wall in a
larger universe was discussed in V.A. Rubakov and M.E. Shaposhnikov,
\textit{Phys. Lett.} {\bf 125B}, 136, 1983.
\bibitem{akama}Another early domain-wall scenario is described in
K. Akama, in {\em Lecture Notes in Physics, 176, Gauge Theory and 
Gravitation, Proceedings, }Nara, 1982;
K. Kikawa, N. Nakanishi, and H. Nariai, Eds.;
Springer-Verlag: Berlin, 1983, p 267, hep-th/0001113.
\bibitem{otherdomain} String inspired domain wall universes are 
discussed in, e.g., A. Lukas et. al., 
\textit{Phys. Rev. D} {\bf 59}, 086001, 1999;
L. Randall and R. Sundrum, \textit{Phys. Rev. Lett.} {\bf 83}, 4690, 1999.
\bibitem{expand} A. Reiss et. al.,  
\textit{Astron. J.}, {\bf 116}, 1009, 1998;
S. Perlmutter et. al., 
\textit{Nature}, {\bf 39}, 51, 1998; P.M. Garnavich et. al.,
\textit{Ap. J.,} {\bf 493}, L53, 1998.
\bibitem{dcp}P.C. Hohenberg and B.I. Halperin, \textit{Rev. Mod. Phys.} 
{\bf 49}, 435, 1977; J. Zinn-Justin, {\em Quantum Field Theory
and Critical Phenomena}, 3rd ed.; 
Clarendon Press: Oxford, 1996, pp 764-777.
\bibitem{nelson}E. Nelson, {\em Quantum Fluctuations}; Princeton
University Press: Princeton, 1985; E. Nelson, \textit{Phys. Rev. }{\bf 150}, 
1079, 1966; F. Guerra, \textit{Phys. Rep.} {\bf 77}, 263, 1981;
Ph. Blanchard, Ph. Comb and W. Zheng, {\em Mathematical and Physical
Aspects of Stochastic Mechanics}; Springer-Verlag: Berlin, 1987.
\bibitem{parisiwu} G. Parisi and Y. Wu, \textit{Scientia Sinica} 
{\bf 24}, 483, 1981.
\bibitem{minkpi} H. H\"{u}ffel and H. Rumpf, \textit{Phys. Lett.} \textbf{148B}, 104, 1984. 
\bibitem{othermink}E. Gozzi, \textit{Phys. Lett. }{\bf 150B}, 119, 1985;
H. Nakazato and Y. Yamanaka, \textit{Phys. Rev. D} {\bf 34}, 492, 1986;
H. Nakazato, \textit{Prog. Theor. Phys.} {\bf 77}, 20, 1987.
\bibitem{namiki}M. Namiki and S. Tanaka, in {\em Modern Problems of
Theoretical Physics - Feschrift for Professor D. Ivanenko};
P.I. Pronin and Yu.N. Obukhov, Eds.; World Scientific: Singapore, 1991.
\bibitem{haken}H. Haken {\em Synergetics}; Springer-Verlag: Berlin,
1977, especially Ch. 6-8; \textit{Rev. Mod. Phys.} {\bf 47}, 67, 1975.
\bibitem{dekker} H. Dekker, Physica {\bf 95A}, 311, 1979; \textit{Phys. Rep.}
{\bf 80}, 1, 1981.
\bibitem{laser}H. Haken in {\em Encyclopedia of Physics, 
vol. XXV/2c: Laser Theory}; S. Fl\"{u}gge, Ed.; 
Springer-Verlag: Berlin, 1970.
\bibitem{frenkel}J. Frenkel and T. Kontorowa, 
\textit{Phys. Z. Sowjet} {\bf13}, 1 (1938). 
\bibitem{frank} F.C. Frank, 
\textit{Proc. Phys. Soc.} {\bf A62}, 131, 1949;
J. Eshelby, \textit{Proc. Phys. Soc.} {\bf A62}, \linebreak 307, 1949.
\bibitem{weertman}J. Weertman and J.R. Weertman, in {\em Dislocations
in Solids - vol. 3: Moving Dislocations}; F.R.N. Nabarro, Ed.;
North-Holland Publishing Co.: Amsterdam, 1980.
\bibitem{nabarro}F.R.N. Nabarro, {\em Theory of Crystal Dislocations};
Oxford University Press: London, 1967.
\bibitem{unziker} This issue is addressed 
in A. Unzicker, gr-qc/0011064.
\bibitem{gunther} H. G\"{u}nther, \textit{Phys. Stat. Sol.} {\bf b149},
101, 1988; H. G\"{u}nther in {\em 
Proceedings of the Eighth International Symposium on
Continuum Models and Discrete
Systems, }Varna, Bulgaria, 1995; K.Z. Markov,Ed.;
World Scientific: Singapore, 1996, p 507.
\bibitem{miller} A.I. Miller, in {\em Some Strangeness in the 
Proportion}; H. Woolf, Ed.; Addison-Wesley: Reading MA,
1980, p 66.
\bibitem{catanese}M. Catanese and T.C. Weeks, \textit{Publ. Astron. Soc. 
Pac.} {\bf 111}, 1193, 1999; B. Degrange and M. Punch, 
\textit{C.R. Acad. Sci. Paris,} {\bf 1}, 189, 2000.
\bibitem{cosmicray}A.V. Olinto, \textit{Phys. Rep.} {\bf 333-334}, 329, 2000;
A.A. Watson,\textit{ Phys. Rep. }{\bf 333-334}, 309, 2000; M. Boratav and
A.A. Watson, \textit{C.R. Acad. Sci. Paris}, {\bf 1}, 207, 2000.
\bibitem{gzk}K. Greisen, \textit{Phys. Rev. Lett.} {\bf 16}, 748, 1966;
G.T. Zatsepin and V.A. Kuzmin, \textit{Sov. Phys. JETP Lett.} {\bf 4}, 78, 1966.
\bibitem{ac}G. Amelino-Camelia and T. Piran, \textit{Phys. Rev. D} {\bf 64},
036005, 2001  and references therein.
\bibitem{farrar}G.R. Farrar and P.L. Biermann, \textit{Phys. Rev. Lett.}
{\bf 81}, 3579, 1998.
\bibitem{rayleigh}Lord Rayleigh (J.W. Strutt), \textit{London Math. Soc. Proc.} 
{\bf 17}, 4, 1885, available in {\em Scientific Papers of
Lord Rayleigh, vol II}; Dover: New York, 1964, p 441.
\bibitem{saw}G.W. Farnell, in {\em Physical Acoustics vol. VI};
Academic Press: New York, 1970, p 109; L.D. Landau and E.M. Lifshitz,
{\em Theory of Elasticity}; Pergamon Press: London, 1959, p 105;
A.A. Maradudin in {\em Nonequilibrium Phonon Dynamics}; 
W.E. Bron, Ed.; Plenum Press: New York, 1985, p 395.
\bibitem{latdyn}I.M. Lifshitz and A.M. Kosevich, in 
{\em Lattice Dynamics}; W.A. Benjamin: New York, 1969, p 53.
\bibitem{viktorov}I.A. Viktorov, \textit{Sov. Phys. Acoustics}, 
{\bf 25}, 1, 1979.
\bibitem{maugin}G.A. Maugin in {\em Recent Developments in
Surface Acoustic Waves}; Springer-Veralg: 
Berlin, 1988, p. 158.
\bibitem{murdoch} A.I. Murdoch, \textit{J. Mech. Phys. Solids},
{\bf 24}, 137, 1976.
\bibitem{farnell}G.W. Farnell and E.L. Adler in {\em Physical
Acoustics, Vol. IX}; W.P. Mason and R.N. Thurston, Eds.; 
Academic Press: New York, 1972.
\bibitem{love}A.E.H. Love, {\em Some Problems of Geodynamics};
Cambridge University Press: London, 1911.
\bibitem{tiersten}H.F. Tiersten, \textit{J. Appl. Phys.} 
{\bf 40}, 770, 1969; H.F. Tiersten, B.K. Sinha, and 
T.R. Mecker, \textit{J. Appl. Phys.} {\bf 52}, 5614, 1981.
\bibitem{kosevich} A.M. Kosevich and A.V. Tutov in {\em Continuum
Models and Discrete Systems - Proceedings of the 8th International
Symposium, }June 11-16, 1995, Varna, Bulgaria; World Scientific: 
Singapore, 1996, p 444.
\bibitem{vpower} N. Daher and G.A. Maugin, \textit{Acta Mech.} {\bf 60}, 
217, 1986. For additional information on excitations on
moving interfaces see
W. Kosi\'{n}ski, {\em Field Singularities and Wave Analysis
in Continuum Mechanics}; Wiley: New York, 1986, 
and references therein.
\bibitem{cottham}M.G. Cottam and D.R. Tilley, 
{\em Introduction to Surface and Superlattice Excitations};
Cambridge University Press: Cambridge, 1989.
\bibitem{chiral}D.B. Kaplan, \textit{Phys. Lett. B} {\bf 288}, 342, 1992;
M. Creutz, \textit{Rev. Mod. Phys.} {\bf 73}, 119, 2001, and references therein.
\bibitem{ash}N.W. Ashcroft and N.D. Mermin, {\em Solid State Physics};
Saunders College Publishing: Philadelphia, 1976, pp 784-789.
\bibitem{fermi}E. Fermi, J. Pasta and S. Ulam, in {\em The 
Collected Papers of Enrico Fermi, vol. 2}; University
of Chicago Press: Chicago, 1966, p 977 (original report
dated 1955).
\bibitem{rasetti} M. Rasetti, {\em Modern Methods in Equilibrium 
Statistical Mechanics}; World Scientific: Singapore, 1986.
\bibitem{robertson} H.S. Robertson, {\em Statistical Thermophysics};
Prentice Hall: Englewood Cliffs, NJ, 1993, pp 43-48. 
\bibitem{ford} J. Ford and G.H. Lunsford, \textit{Phys. Rev. A} 
{\bf 1}, 59, 1970.
\bibitem{anderson}P.W. Anderson, {\em Basic Notions of Condensed 
Matter Physics}; Addison-Wesley: Reading, MA, 1984, pp 30-69.
\bibitem{neeman} Y. Ne'eman, \textit{Proc. Natl. Acad. Sci. USA} {\bf 80},
7051, 1983;\textit{ Found. Phys.} {\bf 16}, 361, 1986.
\bibitem{zim}G.T. Zim\'{a}nyi and K. Vlad\'{a}r, 
\textit{Phys. Rev. A} {\bf 34}, 3496, 1986;  
\textit{Found. Phys. Lett.} {\bf 1}, \linebreak 175, 1988.
\bibitem{ssb}M. Grady, hep-th/9409049.
\bibitem{burton}C.V. Burton, London, Edinburgh, and 
Dublin \textit{Philosophical
Magazine and Journal of Science} {\bf 33}, 191, 1892.
\bibitem{larmor} J. Larmor, {\em Aether and Matter}; Cambridge
University Press: Cambridge, 1900; J. Larmor,
\textit{Phil. Trans. Roy. Soc. London} {\bf A190}, 205, 1897.
\bibitem{ether} K.F. Schaffner, {\em Nineteenth Century Aether
Theories}; Pergamon Press: Oxford, 1972.
\bibitem{eshelby}J.D. Eshelby, \textit{Phys. Rev.} {\bf 90}, 248, 1953.
\bibitem{dislocations}J.P. Hirth and J. Lothe, {\em Theory of
Dislocations}; McGraw Hill: New York, 1968.
\bibitem{dewitt} R. de Wit, in {\em Solid State Physics, vol. 10};
F. Sietz and D. Turnbull, Eds.; Academic Press: New York, 
1960, p 249.
\bibitem{kroner}E. Kr\"{o}ner, \textit{Int. J. Theor. Phys.} {\bf 29}, 
1219, 1990; E. Kr\"{o}ner in {\em Les Houches XXXV - Physics
of Defects};  R. Balian, M. Kl'{e}man and
 J-P. Poirier, Eds.; North-Holland Publishing Company: Amsterdam,
1981, p 215.
\bibitem{sakharov}A.I. Sakharov, \textit{Dokl. Akad. Nauk SSSR, }{\bf 177},
70, 1967 (\textit{Sov. Phys. Doklady }{\bf 12}, \linebreak 1040, 1968).
\bibitem{kokarev}S.S. Kokarev, \textit{Nuovo Cim.} {\bf 113B}, 1339, 1998;  
\textit{Nuovo Cim.} {\bf 114B}, 903, 1999.
\bibitem{3dg} See, e.g. C. Malyshev, \textit{Ann. Phys. }
{\bf 286}, 249, 2000, and references therein; also numerous 
articles in
{\em RAAG Memoirs of the Unifying Study of Basic
Problems in Engineering and Physical Sciences by Means of 
Geometry, vol. I-IV}; K. Kondo, Ed.;  Gakujutsu Bunken Fukyu-Kai: Tokyo, 1968.
\bibitem{vilenkin}A. Vilenkin and E.P.S. Shellard, {\em Cosmic
Strings and Other Topological Defects}; Cambridge University Press:
Cambridge, 1994.
\bibitem{extram}A. Trzesowski, Int. J. \textit{Theor. Phys.} {\bf 33}, 
931, 1994.
\bibitem{hehl}F.W. Hehl, P. von der Hyde and G.D. Kerlick,
\textit{Rev. Mod. Phys.} {\bf 48}, 393, 1976; F.W. Hehl, J.D. McCrea,
E.W. Mielke, and Y. Ne'eman, \textit{Phys. Rep. }{\bf 258}, 1, 1995.
\bibitem{gokeler} M. G\"{o}ckeler and T. Sch\"{u}cker, {\em 
Differential
Geometry, Gauge Theories, and Gravity}; Cambridge University Press:
Cambridge, 1987, ch. 5.
\bibitem{ekp}J. Khoury, B.A. Ovrut, P.J. Steinhardt, and N. Turok,
\textit{Phys. Rev. D }{\bf 64}, 123522, 2001.
\bibitem{anderson2}P.W. Anderson, \textit{Phys. Rev. }{\bf 130}, 439, 1963.
\bibitem{volovik} G.E. Volovik, in {\em Topological Defects
and Non-Equilibrium Dynamics of Symmetry Breaking Phase Transitions};
Y.M. Bunkov and H. Godfrin, Eds.; Kluwer Academic Publishers: 
Dordrecht, 2000, pp 353-387, cond-mat/9902171.
\bibitem{edelen}A. Kadi\'{c} and D.G.B. Edelen, {\em A Gauge Theory
of Dislocations and Disclinations}; Springer-Verlag: Berlin, 1983.
\bibitem{kleinert} H. Kleinert, {\em Gauge Fields in Condensed 
Matter, Vol. II - Stresses and Defects}; World Scientific: Singapore,
1989.
\bibitem{lorentzviol}S. Coleman and S.L. Glashow, \textit{Phys. Rev. D}
{\bf 59}, 116008, 1999; D. Colladay and V.A. Kosteleck\'{y},
\textit{Phys. Rev. D} {\bf 55}, 6760, 1997; {\bf 58}, 16002, 1998.
\bibitem{hoffman} B. Hoffman (with H. Dukas), 
{\em Albert Einstein: Creator and Rebel}; Viking: New York, 1972, p 258.
\bibitem{davies}P.C.W. Davies, {\em The Physics of Time Asymmetry};
University of California Press: Berkely, 1976, p 21.
\bibitem{whitehead} A. N. Whitehead, {\em Process and Reality};
corrected edition, D.R. Griffin and D.W. Sherburne Eds.;
Free Press: New York, 1978; {\em Physics and the Ultimate
Significance of Time}; D. R. Griffin, Ed.; SUNY Press: Albany, 1986.
\end{thebibliography}
\end{document}